\documentclass[twocolumn,trackchanges]{aastex701}

\usepackage[nolist]{acronym}
\usepackage{xcolor}
\usepackage{comment}
\usepackage{xspace}
\usepackage{hyperref}

\definecolor{darkblue}{HTML}{00235f}

\DeclareRobustCommand{\ion}[2]{\textup{#1\,\sc{\lowercase{#2}}}}
\newcommand*\element[1][]{%
  \def\aa@element@tr{#1}%
  \aa@element
}

\newcommand*{\eg}{e.g.\@\xspace}

\newcommand{\Mm}{~\mathrm{Mm}}
\newcommand{\G}{~\mathrm{G}}

\newcommand{\km}{~\mathrm{km}}

\newcommand{\kms}{\ensuremath{\, \mathrm{km\,s^{-1}} }\xspace}
\newcommand{\ppm}{\ensuremath{\, \mathrm{pm} }\xspace}

\newcommand{\mghk}{\ion{Mg}{II} h\&k\xspace}

\newcommand{\snapfiveOneEight}{\texttt{muram\_en\_518000\_503s}\xspace}

\newcommand{\caefft}{\ion{Ca}{ii}~$\lambda854.2$~nm\xspace}

\begin{document}

\begin{acronym}

\acro{lte}[LTE]{local-thermodynamic-equilibrium}
\acro{rte}[RTE]{radiative transfer equation}
\acro{te}[TE]{thermodynamic equilibrium}
\acro{ne}[NE]{nonequilibrium}

\acro{nlte}[NLTE]{non-local-thermodynamic-equilibrium}
\acro{se}[SE]{statistical equilibrium}
\acro{los}[LOS]{line-of-sight}
\acro{eos}[EoS]{equation of state}

\acro{crd}[CRD]{complete frequency redistribution}
\acro{prd}[PRD]{partial frequency redistribution}
\acro{rh15d}[RH1.5D]{Rybicki \& Hummer 1.5D RT code}
\acro{rt}[RT]{radiative transfer}
\acro{mhd}[MHD]{magnetohydrodynamics}
\acro{rmhd}[rMHD]{radiative-magnetohydrodynamics}
\acro{iris}[IRIS]{Interface Region Imaging Spectrometer }
\acro{muram}[MURaM]{Max Planck Institute for Solar System Research/University of Chicago Radiation Magneto-hydrodynamics}
\acro{muramche}[MURaM-ChE]{chromospheric extension of MURaM}
\acro{clv}[CLV]{center-to-limb variation}
\acro{euv}[EUV]{extreme ultra violet}
\acro{nir}[NIR]{near-infrared}
\acro{nuv}[NUV]{near-ultraviolet}
\acro{uv}[UV]{ultraviolet}
\acro{ali}[ALI]{approximate Lambda iteration}
\acro{ff}[ff]{free-free}
\acro{qs}[QS]{quiet Sun}
\acro{soup}[SOUP]{Solar Optical Universal Polimeter}
\acro{sst}[SST]{Swedish $1 \, \mathrm{m}$ Solar Telescope}
\acro{chromis}[CHROMIS]{CHROMospheric Imaging Spectrometer}
\acro{gong}[GONG]{Global Oscillation Network Group}
\acro{dot}[DOT]{Dutch open telescope}
\acro{ar}[AR]{active regions}
\acro{fb}[fb]{free-bound}
\acro{bf}[bf]{bound-free}
\acro{en}[EN]{enhanced network}
\acro{fov}[FOV]{field-of-view}

\acro{bb}[bb]{bound-bound}
\acro{fwhm}[FWHM]{full width at half maximum}
\acro{ssd}[SSD]{small-scale dynamo}
\acro{fts}[FTS]{Fourier Transform Spectrograph}

\acro{im}[IM]{isotope model}
\acro{sim}[SIM]{single isotope model}
\acro{cm}[CM]{composite model}
\acro{ftsatlas}[Hamburg FTS atlas]{Hamburg Fourier-transform-spectrograph atlas}

\acro{sfp}[SFPs]{strong-field-profiles}
\acro{wfp}[WFPs]{weak-field-profiles}

\acro{ufp}[UFPs]{upflow-profiles}

\acro{dfp}[DFPs]{downflow-profiles}

\acro{opi}[OPI]{opposite-polarity-intrusion}
\acro{tfr}[TFR]{twisted flux rope}
\acro{tcp}[TCP]{total circular polarization}

\acro{scip}[SCIP]{Sunrise Chromospheric Infrared spectroPolarimeter}
\acro{hmi}[HMI]{Helioseismic Magnetic Imager}
\acro{phi}[PHI]{Polarimetric and Helioseismic Imager}
\acro{rre}[RRE]{rapid redshifted exursion}
\acro{rbe}[RBE]{rapid blueshifted exursion}

\end{acronym}

\title{Cause of chromospheric opposite polarity intrusions discovered in Sunrise III/SCIP data: MURaM-ChE simulations point to twisted flux ropes}

\author[orcid=0000-0003-2038-6062,sname='Ondratschek']{Patrick~A.~Ondratschek} \affiliation{Max-Planck-Institut für Sonnensystemforschung, Justus-von-Liebig-Weg 3, 37077 Göttingen, Germany}\email{ondratschek@mps.mpg.de}

\author[orcid=0000-0003-1670-5913,sname='Przybylski']{Damien F. Przybylski} \affiliation{Max-Planck-Institut für Sonnensystemforschung, Justus-von-Liebig-Weg 3, 37077 Göttingen, Germany}\email{przybylski@mps.mpg.de}	

\author[sname='Cameron']{Robert H. Cameron} \affiliation{Max-Planck-Institut für Sonnensystemforschung, Justus-von-Liebig-Weg 3, 37077 Göttingen, Germany}\email{cameron@mps.mpg.de}	

\author[orcid=0000-0003-3490-6532,sname='Smitha']{H.~N.~Smitha} \affiliation{Max-Planck-Institut für Sonnensystemforschung, Justus-von-Liebig-Weg 3, 37077 Göttingen, Germany}\email{narayanamurthy@mps.mpg.de}		

\author[orcid=0000-0002-3418-8449,sname='Solanki']{Sami~K.~Solanki} \affiliation{Max-Planck-Institut für Sonnensystemforschung, Justus-von-Liebig-Weg 3, 37077 Göttingen, Germany}\email{solanki@mps.mpg.de}		

\author[orcid=0000-0001-5616-2808,sname='Kubo']{Masahito~Kubo} \affiliation{National Astronomical Observatory of Japan, 2-21-1 Osawa, Mitaka, Tokyo 181-8588, Japan}\email{masahito.kubo@nao.ac.jp}		

\author[orcid=0000-0003-1459-7074,sname='Lagg']{Andreas~Lagg} \affiliation{Max-Planck-Institut für Sonnensystemforschung, Justus-von-Liebig-Weg 3, 37077 Göttingen, Germany}\email{lagg@mps.mpg.de}		
\author[orcid=0000-0002-9972-9840,sname='Gandorfer']{Achim~Gandorfer} \affiliation{Max-Planck-Institut für Sonnensystemforschung, Justus-von-Liebig-Weg 3, 37077 Göttingen, Germany}\email{gandorfer@mps.mpg.de}		
\author[orcid=0000-0002-3387-026X,sname='del~Toro~Iniesta']{Jose~Carlos~del~Toro~Iniesta} \affiliation{Instituto de Astrofísica de Andalucía, CSIC, Glorieta de la Astronomía s/n, 18008 Granada, Spain}\affiliation{Spanish Space Solar Physics Consortium}\email{jti@iaa.es}		
\author[orcid=0000-0002-5054-8782,sname='Katsukawa']{Yukio~Katsukawa} \affiliation{National Astronomical Observatory of Japan, 2-21-1 Osawa, Mitaka, Tokyo 181-8588, Japan}\affiliation{Department of Astronomy, The University of Tokyo, 7-3-1, Hongo, Bunkyo-ku, Tokyo 113-0033, Japan}\affiliation{Department of Astronomical Science, The Graduate University for Advanced Studies (SOKENDAI), 2-21-1 Osawa, Mitaka, Tokyo 1818588, Japan}\email{yukio.katsukawa@nao.ac.jp}		
\author[orcid=0000-0002-0787-8954,sname='Bernasconi']{Pietro~Bernasconi} \affiliation{Johns Hopkins University Applied Physics Laboratory, 11100 Johns Hopkins Road, Laurel, Maryland, USA}\email{pietro.bernasconi@jhuapl.edu}		
\author[sname='Berkefeld']{Thomas~Berkefeld} \affiliation{Institut für Sonnenphysik (KIS), Georges-Köhler-Allee 401a, 79110 Freiburg, Germany}\email{thomas.berkefeld@leibniz-kis.de}		
\author[orcid=0009-0009-4425-599X,sname='Feller']{Alex~Feller} \affiliation{Max-Planck-Institut für Sonnensystemforschung, Justus-von-Liebig-Weg 3, 37077 Göttingen, Germany}\email{feller@mps.mpg.de}		
\author[orcid=0000-0001-6317-4380,sname='Riethmüller']{Tino~L.~Riethmüller} \affiliation{Max-Planck-Institut für Sonnensystemforschung, Justus-von-Liebig-Weg 3, 37077 Göttingen, Germany}\email{riethmueller@mps.mpg.de}

\author[orcid=0000-0001-9228-3412,sname='Álvarez-Herrero']{Alberto~Álvarez-Herrero} \affiliation{Instituto Nacional de T\'ecnica Aeroespacial (INTA), Ctra. de Ajalvir, km. 4, E-28850 Torrejón de Ardoz, Spain}\affiliation{Spanish Space Solar Physics Consortium}\email{alvareza@inta.es}		
	
\author[orcid=0000-0001-8829-1938,sname='Orozco~Suárez']{David~Orozco~Suárez} \affiliation{Instituto de Astrofísica de Andalucía, CSIC, Glorieta de la Astronomía s/n, 18008 Granada, Spain}\affiliation{Spanish Space Solar Physics Consortium}\email{orozco@iaa.es}		
\author[sname='Grauf']{Bianca~Grauf} \affiliation{Max-Planck-Institut für Sonnensystemforschung, Justus-von-Liebig-Weg 3, 37077 Göttingen, Germany}\email{grauf@mps.mpg.de}		
\author[sname='Carpenter']{Michael~Carpenter} \affiliation{Johns Hopkins University Applied Physics Laboratory, 11100 Johns Hopkins Road, Laurel, Maryland, USA}\email{michael.carpenter@jhuapl.edu}		
\author[sname='Bell']{Alexander~Bell} \affiliation{Institut für Sonnenphysik (KIS), Georges-Köhler-Allee 401a, 79110 Freiburg, Germany}\email{albe@leibniz-kis.de}		
\author[orcid=0000-0001-7764-6895,sname='Martínez~Pillet']{Valentín~Martínez~Pillet} \affiliation{Instituto de Astrofísica de Canarias, Vía Láctea, s/n, E-38205 La Laguna, Spain}\affiliation{Spanish Space Solar Physics Consortium}\email{vmpillet@iac.es}		
\author[orcid=0000-0001-7696-8665,sname='Gizon']{Laurent~Gizon} \affiliation{Max-Planck-Institut für Sonnensystemforschung, Justus-von-Liebig-Weg 3, 37077 Göttingen, Germany}\affiliation{Institut für Astrophysik und Geophysik, Georg-August-Universität Göttingen, 37077 Gōttingen, Germany}\email{gizon@mps.mpg.de}

\author[orcid=0000-0002-7318-3536,sname='Bailén']{Francisco~Javier~Bailén} \affiliation{Instituto de Astrofísica de Andalucía, CSIC, Glorieta de la Astronomía s/n, 18008 Granada, Spain}\affiliation{Spanish Space Solar Physics Consortium}\email{fbailen@iaa.es}		
\author[orcid=0000-0002-2055-441X,sname='Blanco~Rodríguez']{Julian~Blanco~Rodríguez} \affiliation{Universitat de Valencia Catedrático José Beltrán 2, E-46980 Paterna-Valencia, Spain}\affiliation{Spanish Space Solar Physics Consortium}\email{julian.blanco@uv.es}		
\author[orcid=0000-0003-4319-2009,sname='Castellanos~Durán']{Juan~Sebastián~Castellanos~Durán} \affiliation{Max-Planck-Institut für Sonnensystemforschung, Justus-von-Liebig-Weg 3, 37077 Göttingen, Germany}\email{castellanos@mps.mpg.de}		
\author[orcid=0009-0002-6808-5154,sname='Harnes']{Edvarda~Harnes} \affiliation{Max-Planck-Institut für Sonnensystemforschung, Justus-von-Liebig-Weg 3, 37077 Göttingen, Germany}\email{harnes@mps.mpg.de}		
\author[orcid=0000-0001-6029-7529,sname='Hoelken']{Johannes~Hoelken} \affiliation{Max-Planck-Institut für Sonnensystemforschung, Justus-von-Liebig-Weg 3, 37077 Göttingen, Germany}\email{hoelken@mps.mpg.de}		
\author[orcid=0000-0003-1409-1145,sname='Iglesias']{Francisco~A.~Iglesias} \affiliation{Max-Planck-Institut für Sonnensystemforschung, Justus-von-Liebig-Weg 3, 37077 Göttingen, Germany}\affiliation{Grupo de Estudios en Heliofísica de Mendoza, CONICET, Universidad de Mendoza, Boulogne sur Mer 683, 5500 Mendoza, Argentina}\email{iglesias@mps.mpg.de}		
\author[orcid=0000-0002-4669-5376,sname='Ishikawa']{Ryohtaroh~T.~Ishikawa} \affiliation{National Institute for Fusion Science, 322-6 Oroshi-cho, Toki City 509-5292, Japan}\email{ishikawa.ryohtaro@nifs.ac.jp}		
\author[orcid=0000-0001-7452-0656,sname='Kawabata']{Yusuke~Kawabata} \affiliation{National Astronomical Observatory of Japan, 2-21-1 Osawa, Mitaka, Tokyo 181-8588, Japan}\email{kawabata.yusuke@nao.ac.jp}		
\author[orcid=0000-0002-1043-9944,sname='Matsumoto']{Takuma~Matsumoto} \affiliation{Centre for Integrated Data Science, Institute for Space-Earth Environmental Research, Nagoya University, Furocho, Chikusa-ku, Nagoya, Aichi 464-8601, Japan}\email{takuma.matsumoto@gmail.com}		
\author[orcid=0000-0002-7044-6281,sname='Oba']{Takayoshi~Oba} \affiliation{Advanced Research Center for Space Science and Technology, Institute of Science and Engineering, Kanazawa University, Kakuma-machi, Kanazawa, Ishikawa 920-1192, Japan}\affiliation{Max-Planck-Institut für Sonnensystemforschung, Justus-von-Liebig-Weg 3, 37077 Göttingen, Germany}\email{oba@mps.mpg.de}		
\author[orcid=0000-0003-0175-6232,sname='Siu-Tapia']{Azaymi~L.~Siu-Tapia} \affiliation{Instituto de Astrofísica de Andalucía, CSIC, Glorieta de la Astronomía s/n, 18008 Granada, Spain}\affiliation{Spanish Space Solar Physics Consortium}\email{siu@iaa.es}		
\author[orcid=0000-0003-1483-4535,sname='Strecker']{Hanna~Strecker} \affiliation{Instituto de Astrofísica de Andalucía, CSIC, Glorieta de la Astronomía s/n, 18008 Granada, Spain}\affiliation{Spanish Space Solar Physics Consortium}\email{streckerh@iaa.es}		
\author[orcid=0000-0003-1971-5551,sname='Vukadinović']{Dušan~Vukadinović} \affiliation{Institut für Physik, Universität Graz, Universitätsplatz 5, 8010 Graz, Austria}\affiliation{Max-Planck-Institut für Sonnensystemforschung, Justus-von-Liebig-Weg 3, 37077 Göttingen, Germany}\email{vukadinovic@mps.mpg.de}


\author[orcid=0000-0001-5686-3081,sname='Hara']{Hirohisa~Hara} \affiliation{National Astronomical Observatory of Japan, 2-21-1 Osawa, Mitaka, Tokyo 181-8588, Japan}\email{hirohisa.hara@nao.ac.jp}		
\author[orcid=0000-0003-4764-6856,sname='Shimizu']{Toshifumi~Shimizu} \affiliation{Department of Earth and Planetary Science, The University of Tokyo, 7-3-1, Hongo, Bunkyo-ku, Tokyo 113-0033, Japan}\affiliation{Institute of Space and Astronautical Science, Japan Aerospace Exploration Agency, 3-1-1, Yoshinodai, Chuo-ku, Sagamihara, Kanagawa 252-5210, Japan}\email{shimizu.toshifumi@isas.jaxa.jp}

\begin{abstract}

The \ac{scip} instrument onboard the balloon-borne  \textsc{Sunrise~iii} observatory provided new high-resolution observations of the solar chromosphere in the \caefft line. 
The Stokes-$V$ signal in magnetic network regions was found to show fine-structured details, which suggests the magnetic field above the network elements does not simply expand as a unipolar feature but displays \acp{opi}. These features appear as elongated structures in Stokes-$V$ observations. In this work, we demonstrate that such features appear ubiquitously in a numerical simulation of the solar chromosphere. We use a simulation that is computed with the recently developed \ac{muramche} and resembles an enhanced network region. We find that \acp{opi} appear ubiquitously in the vertical component of the magnetic field at around $1$ Mm above the surface and are visible in the synthetic Stokes-$V$ signal of the \caefft line. The structures have lengths of $\approx 2 \Mm$ to $7 \Mm$ and widths of approximately $1 \Mm$. The magnetic field configurations associated with the \ac{opi} features appear to belong to \acp{tfr} and are visible for most of the time in the presented $21 \min$ time series. Our results show that the magnetic structure of the chromosphere is more complex than previously thought, with even seemingly simple flux tubes showing embedded twisted fields pointing in the opposite direction. This may help in explaining new high-resolution observations from the \textsc{Sunrise~iii} mission.

\end{abstract}


\section{Introduction} 
The magnetic network in the solar atmosphere is a part of the \ac{qs} and typically forms at the borders of supergranular cells. Within these magnetic flux concentrations, magnetic field strengths of the order of thousands of Gau\ss$\,$ can be found at the surface \citep[for reviews of \ac{qs} magnetic fields see \eg,][]{2009SSRv..144..275D,2019LRSP...16....1B}. Above the surface, the magnetic field concentrations are expected to expand and form a magnetic canopy at the transition from the photosphere to the chromosphere \citep[][]{1990A&A...234..519S}.

The \ac{los} component of the magnetic field in the photosphere can be inferred from instruments such as the \ac{hmi} \citep[][]{2012SoPh..275..207S} onboard the Solar Dynamics Observatory (SDO) or the \ac{phi} \citep[][]{2020A&A...642A..11S} onboard Solar Orbiter. In the chromosphere, magnetic fields can be inferred for example with the CRISP instrument \citep{2008ApJ...689L..69S} which is mounted onto the \ac{sst} \citep[][]{2003SPIE.4853..341S}. High-resolution and high-precision chromospheric magnetograms are, however, rare. The \ac{scip} instrument \citep[][]{2026SoPh..301...99K} onboard the \textsc{Sunrise~iii}  observatory \citep{2025SoPh..300...75K} provides full Stokes observations of the \caefft line at a resolution of $0.21''$.  \textsc{Sunrise~iii} is the third flight of the Sunrise observatory \citep{2011SoPh..268....1B}. Overviews of the three successful flights can be found in \citet{2010ApJ...723L.127S}, \citet{2017ApJS..229....2S}, and \citet{2026ApJ..1005L..64S}.

\citet{2026ApJ..1008L...9K} present high-resolution observations with \ac{scip} from  \textsc{Sunrise~iii} of the \ac{qs}. The authors derived \ac{los} magnetograms from the Stokes-$V$ signal of the \caefft line that show \acfp{opi} in the presence of otherwise unipolar network magnetic fields. These features appear as elongated structures in the magnetogram. The aim of this work is to identify whether  similar fine-structured magnetic network fields can be found in a numerical model of the solar atmosphere, and to identify 
the underlying magnetic field configuration.

\section{Simulation and forward modeling}
We employ an \ac{en} model computed with the recently developed \ac{muramche} code \citep{2022A&A...664A..91P}. The simulation domain extends $24 \Mm \times 24 \Mm \times 24 \Mm$, including a roughly $7 \Mm$ deep convection zone and a $17 \Mm$ high atmosphere. The horizontal resolution is $23.4 \km$ and the vertical resolution is $20 \km$. For the setup of the simulation, a bipolar magnetic feature of $\approx 8 \Mm$ separation was added on top of a \ac{ssd} simulation \citep[][]{2025A&A...703A.148P}. For a detailed description of the physics included in the model and the setup, we refer to \citet{2022A&A...664A..91P} and \citet{2024A&A...692A...6O}.

We use the RH1.5D code \citep{2001ApJ...557..389U,2015A&A...574A...3P} to forward model the \caefft line including polarization effects. In the RH1.5D code, each vertical column in the atmosphere is treated as an individual plane-parallel atmosphere for the solution of the \ac{rte}. For the synthesis, we employ a five-level-plus-continuum model atom. The \caefft transition is computed in the approximation of \ac{crd}, which is sufficient for this line \citep{1989A&A...213..360U}. The synthesized spectral line data are the same as we used in \citet{2026A&A...708A...3O}.
\section{Results}

\begin{figure*}[ht!]
\includegraphics[width=\textwidth,clip]{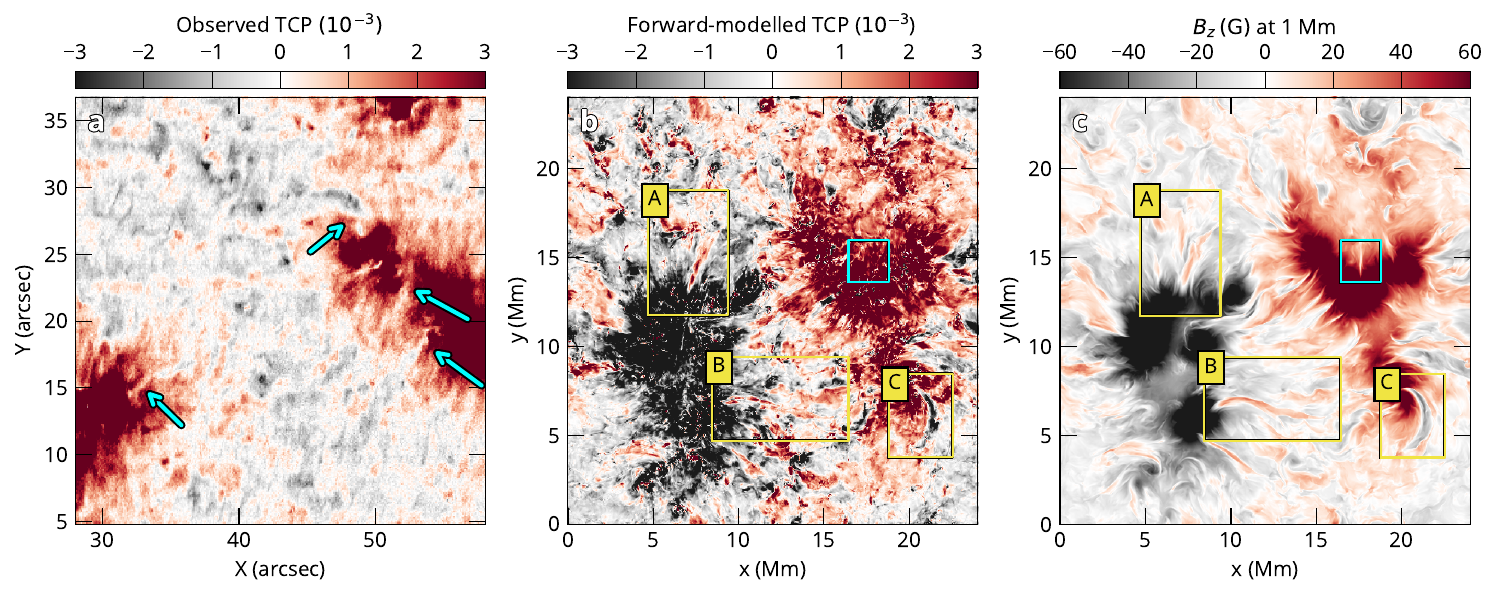}
\caption{Maps of the TCP and the vertical component of the magnetic field. Panel (a) shows a TCP map that is a part of the \ac{qs} dataset observed by \textsc{Sunrise~iii} and discussed in detail in \citet{2026ApJ..1008L...9K}. Panel (b) shows the TCP map from the simulation and panel (c) the vertical magnetic field map at $z=1\Mm$.
Panels (b) and (c) cover the whole horizontal computational domain. The yellow rectangles indicate three locations (A, B, and C) that contain an \ac{opi} feature. The presented data corresponds to snapshot \snapfiveOneEight. The evolution of the vertical magnetic field can be seen in the attached movie in Fig.~\ref{fig:movie_bz}. We note that for an easier comparison, we used the same color table for panels (a), (b), and (c), although they show different quantities. The colormaps are clipped such that the \acp{opi} are clearly visible, while the network magnetic fields are saturated.
\label{fig:fig1-magnetogram-and-stokes-V}}
\end{figure*}

In this section, we describe how the magnetic network fields appear at $1\Mm$ in the simulated atmosphere and in the Stokes-$V$ signal of the forward-modeled \caefft line. Supplementary movies are provided to illustrate the temporal evolution and the 3D topology of the magnetic field.

First, we compare the observed \ac{tcp} map with the forward-modeled \ac{tcp} map. The \ac{tcp} is computed from the Stokes-$V$ signal of the \caefft line and samples the chromosphere. For the detailed calculation of the \ac{tcp} we refer to \citet{2026ApJ..1008L...9K} and App.~\ref{app:tcp}. In Fig.~\ref{fig:fig1-magnetogram-and-stokes-V}, we compare a \ac{qs} observation \citep[for details see,][]{2026ApJ..1008L...9K} to the synthetic \ac{tcp} map and the magnetogram at $z=1\Mm$ for one snapshot of the simulation. The height of $z=1\Mm$ was chosen as it approximately coincides with the formation height of the \caefft line core. We note, however, that the actual formation heights show much more variation \citep[see,][Fig.~1d]{2026A&A...708A...3O}. The \ac{scip} instrument covers a \ac{fov} of $58''\times 58''$ with a pixel scale of $0.094''$. The \ac{fov} shown is a zoom-in to the original observation that is of similar size as the simulation. The pixel scale of the simulation is smaller than in the observation. We have not degraded the forward-modeled \ac{tcp} by any instrumental parameters to match the observations. The observed \ac{tcp} (panel a) clearly shows how the network magnetic fields expand in the chromosphere but show \acp{opi}, as indicated by the light blue arrows at some example positions. The corresponding \ac{tcp} map from the simulation (panel b) shows similar elongated features at the borders of the expanded network magnetic field. In comparison with the vertical magnetic field at $z=1\Mm$ (panel c), regions are found, however, where an elongated \ac{opi} feature is visible in the magnetic field (panel c) but not in the \ac{tcp} map (panel b). An example is given at $(x,y)=(17 \Mm, 15\Mm)$, as indicated by the blue rectangle in panels (b) and (c). This can either be due to a too weak vertical component of the magnetic field or that Stokes-$V$ does not get its main contribution at $1\Mm$ in these particular pixels. Indeed, the whole \ac{tcp} map appears more inhomogeneous than the magnetic field map, probably reflecting the influence of other atmospheric parameters and changes in the height of formation \citep[see \eg, Fig.~1 panels d and f in][for a map of the formation height at the line profile minimum and $B_z$ at the the same height]{2026A&A...708A...3O}.

We highlighted the locations of three features, A, B, and C, in panels (b) and (c) of Fig.~\ref{fig:fig1-magnetogram-and-stokes-V} that show a clear \ac{opi} feature in both, the magnetic field and the corresponding \ac{tcp}. At the locations where the features appear next to the surrounding network polarity, the features are relatively sharply defined. That is, they show clear edges between the main network polarity and the feature of opposite polarity. With increasing distance from the main network polarity, the features appear more diffuse.

\begin{figure*}[ht!]
\includegraphics[width=\textwidth,clip]{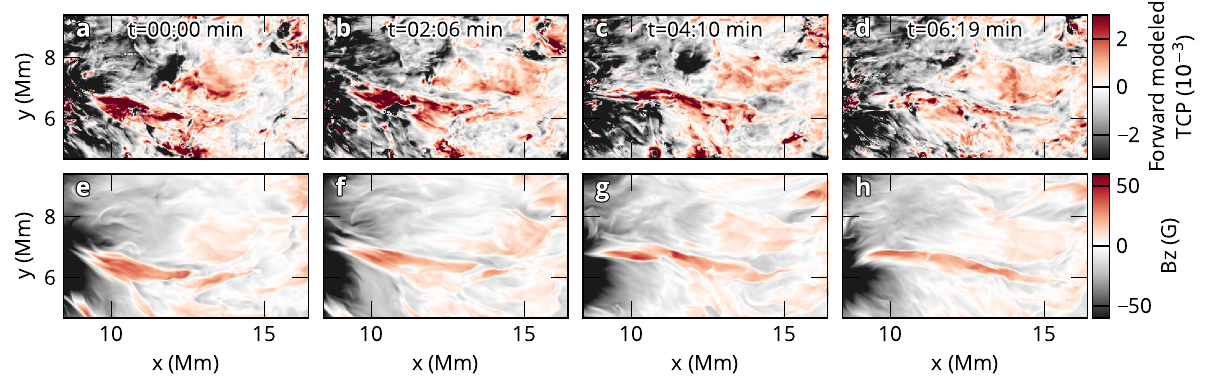}
\caption{Evolution of feature B. In the first row (panels a--d), we show a map of the \ac{tcp} (see Fig.~\ref{fig:fig1-magnetogram-and-stokes-V} as reference). In the bottom row (panels e--h), we show the vertical component of the magnetic field at $1 \Mm$ in the atmosphere at the same region. The two quantities are shown at four different snapshots of the simulation that are separated by approximately $2 \min$ of simulation time. These four snapshots precede the snapshot shown in Fig.~\ref{fig:fig1-magnetogram-and-stokes-V}, which corresponds to $t=8 \min$. 
\label{fig:fig2-feature-B-time-evolution}}
\end{figure*}

\begin{figure*}[ht!]
\begin{interactive}{animation}{animation_figure_3.mp4}
\includegraphics[width=\textwidth,clip]{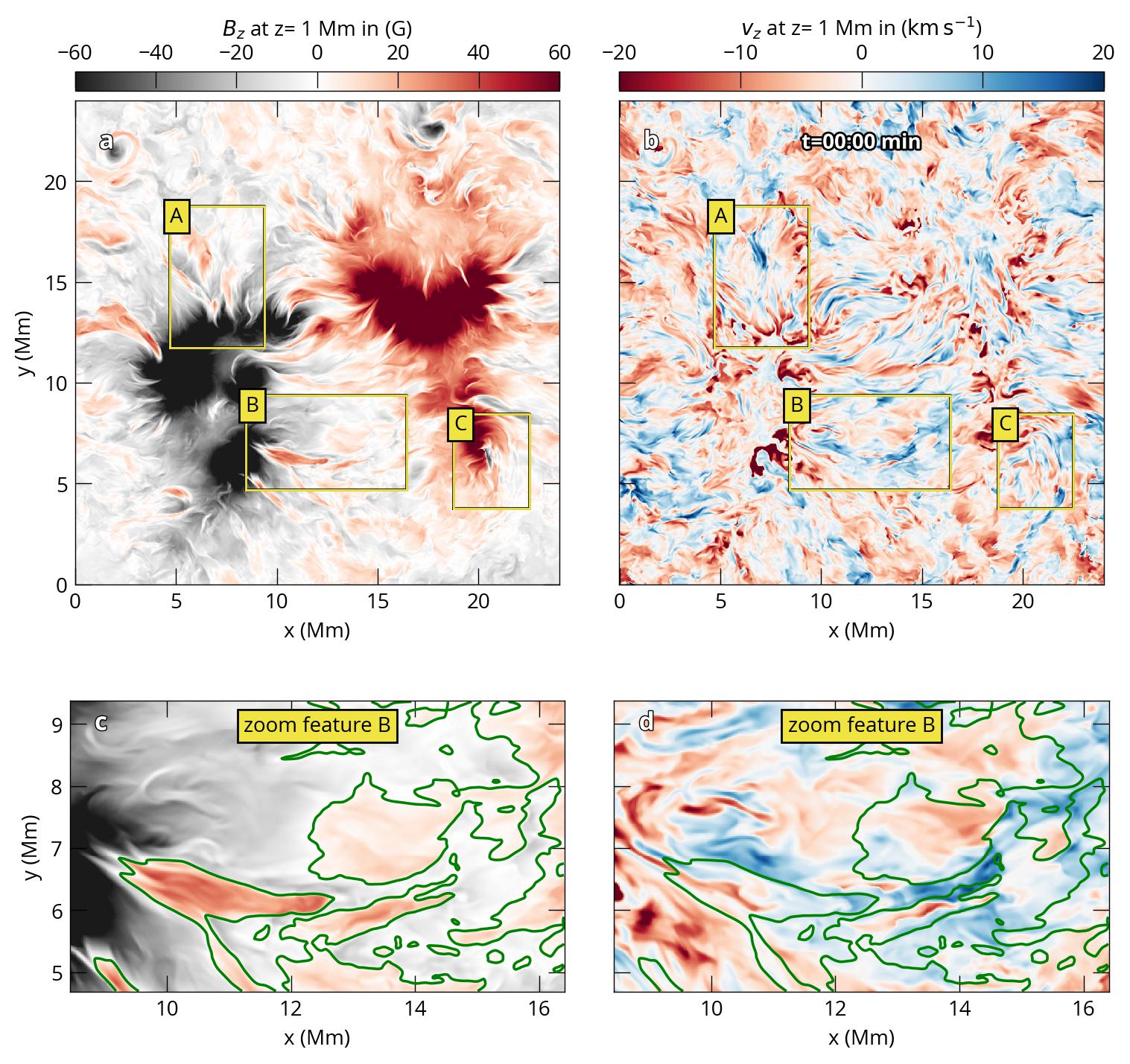}
\end{interactive}\caption{Animation showing the evolution of the vertical magnetic field (panels a and c) and vertical velocity (panels b and d) in the simulated atmosphere at $z=1\Mm$. The movie runs $38 \sec$ and shows $21 \min$ of simulation time. Panels (c and d) display zooms to region B as indicated in panels (a and b). The contours in panels (c and d) enclose regions of positive vertical magnetic field to simplify the comparison with the vertical velocity at the corresponding pixels. An online version of the movie can be found in the \href{https://iopscience.iop.org/article/10.3847/2041-8213/ae9607}{published journal version} of the article.}.
\label{fig:movie_bz}
\end{figure*}

In Fig.~\ref{fig:fig2-feature-B-time-evolution}, we present the evolution of feature B at four timesteps that are separated by $\approx 2 \min$ of simulation time and occur before the snapshot presented in Fig.~\ref{fig:fig1-magnetogram-and-stokes-V}. It can be seen that the feature is visible in the \ac{tcp} maps in each of these snapshots and in the corresponding magnetograms. While the shape and extent varies slightly with time, the overall \ac{opi} structure appears relatively stable over the time series. In the movie attached to Fig.~\ref{fig:movie_bz}, it can be seen from the magnetogram that such structures, for example features A and B, are visible over the whole presented $\approx 21 \min$ time series in the magnetogram at $z=1\Mm$. Some other features, such as feature C, become more visible only towards the end of the time series. We note however, that the movie shows the vertical magnetic field at $z=1 \Mm$. A feature that appears or disappears in the movie might simply move out of one of the presented height ranges and not necessarily appear or disappear in the simulation as a whole.

\begin{figure*}[ht!]
\includegraphics[width=\textwidth,clip]{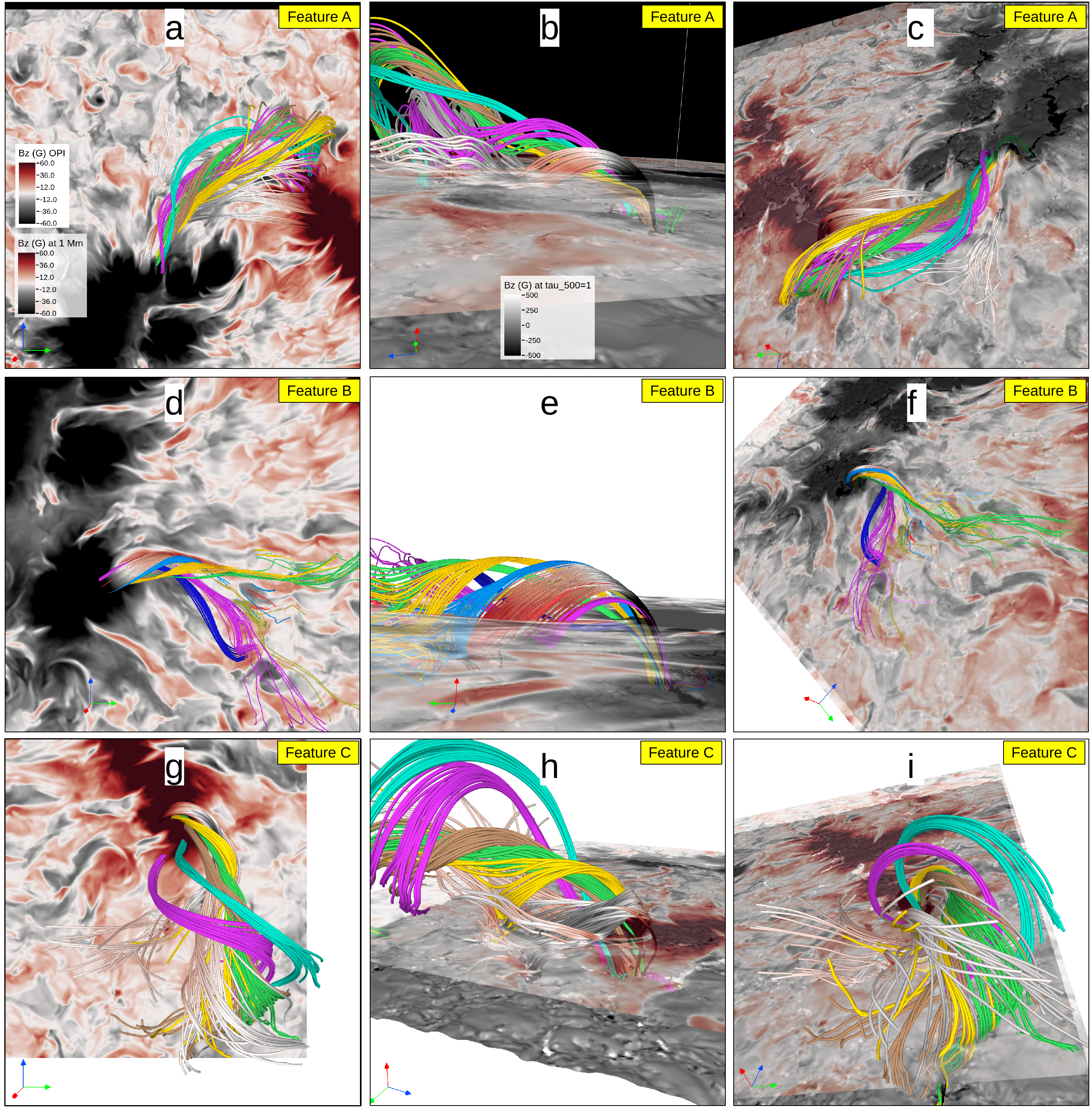}
\caption{3D rendering of magnetic field lines of features A, B, and C. All panels show 3D rendered magnetic field lines from snapshot \snapfiveOneEight. For better visualization, we colored the magnetic field lines associated with the \ac{opi} according to the value of the vertical magnetic field using the same color scale as in Fig.~\ref{fig:fig1-magnetogram-and-stokes-V}. The other magnetic field line sets are shown in a constant color to highlight the twisted magnetic field structure. Panel (a) shows a view from the top similar to Fig.~\ref{fig:fig1-magnetogram-and-stokes-V} with the vertical magnetic field at $1 \Mm$ in the background. In the other panels, the magnetic field at $z=1\Mm$ is shown semi-transparently. Panel (b) shows a side view with a focus on the \ac{opi}, and in the background, the vertical magnetic field at the $\tau_{500}=1$ surface is shown. Panel (c) shows a different viewing angle to highlight the twisted magnetic field structure.  Panels (d)-(e)-(f) and (g)-(h)-(i) are analogous to panels (a)-(b)-(c), but for features B and C, as indicated in the top right corner of the panels. The 3D renderings of the atmosphere and the magnetic field lines were created with Vapor \citep[][]{2019Atmos..10..488L,2023zndo...7779648S}. In App.~\ref{app:animations_of_features_a_b_c} we provide animations that show a $360^{\circ}$ view of features A (Fig.~\ref{fig:animation_feature_a}), B (Fig.~\ref{fig:animation_feature_b}), and C (Fig.~\ref{fig:animation_feature_c}). 
\label{fig:fig3-3d-rendering-of-field-lines}}
\end{figure*}

We present the associated magnetic field of features A, B, and C as 3D-rendered field lines at one snapshot in Fig.~\ref{fig:fig3-3d-rendering-of-field-lines}. In panels (a), (d), and (g), that is in the first column of the figure, we show the vertical magnetic field component at $1 \Mm$ in the background to put the magnetic field lines into the context of Fig.~\ref{fig:fig1-magnetogram-and-stokes-V} (c). The magnetic field lines that are related to the corresponding \ac{opi} are color-coded by the vertical magnetic field component. We show additional sets of magnetic field lines in constant color that are neighboring the \ac{opi} feature. These strands were selected manually to visualize the coiling of the magnetic field lines. In panels (b), (e), and (h), that is in the second column of the figure, the viewpoint is chosen such that the magnetic field belonging to the \ac{opi} is clearly visible from the side. Finally, panels (c), (f), and (i), that is the third column of the figure, present a different viewing angle for all three features to further illustrate the 3D structure of the magnetic field topology. 

\begin{figure*}[ht!]

\begin{interactive}{animation}{animation_figure_5.mp4}
\includegraphics[width=\textwidth,clip]{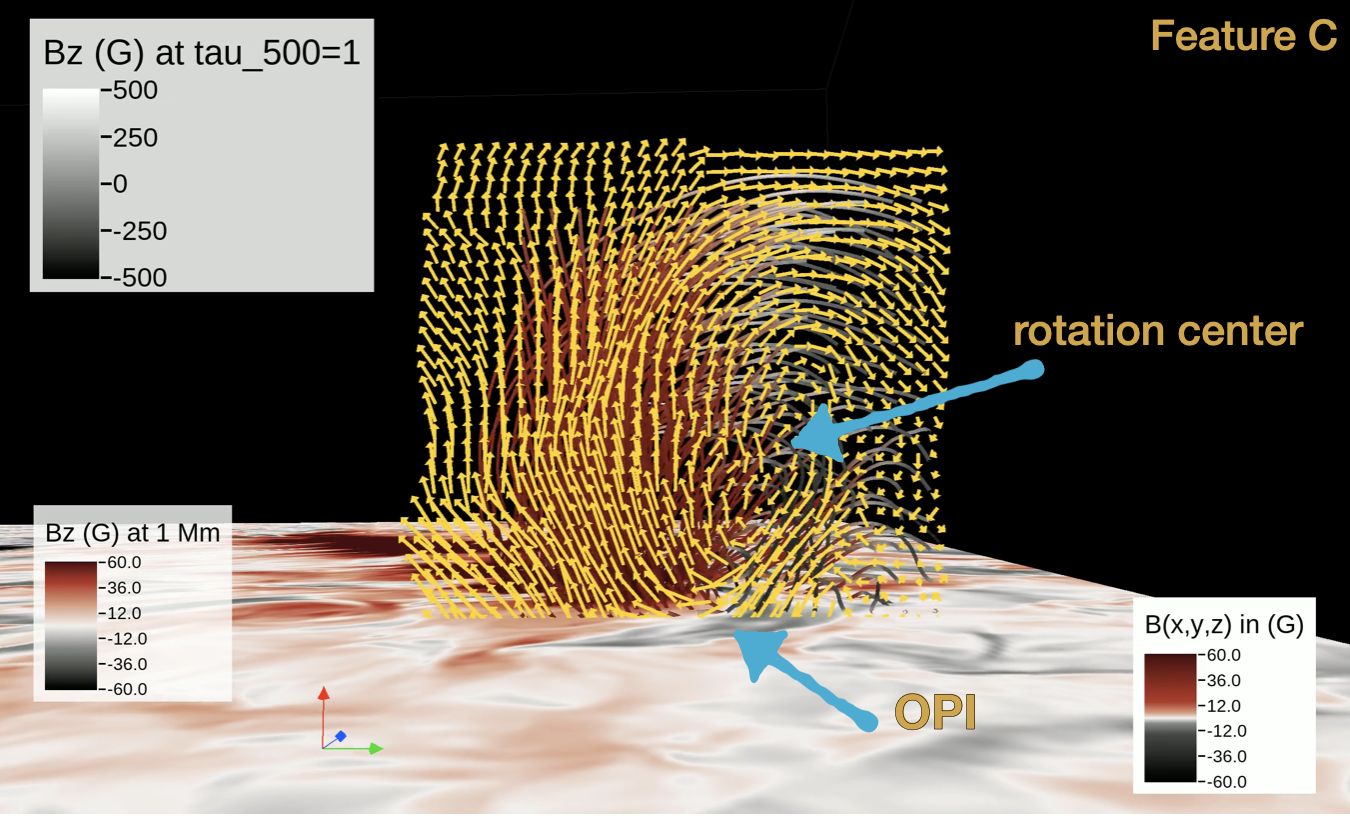}
\end{interactive}

\caption{Animation of a cross section across the twist axis of feature C. The animation shows yellow arrows indicating the direction of the magnetic field in a vertical cut through the \ac{tfr} that is approximately orthogonal to the central rotation axis of the \ac{tfr}. In the background, a horizontal slice indicates the vertical component of the magnetic field at $z=1 \Mm$. The example frame additionally shows an indication of the \ac{opi} defining feature C, and the approximate position of the coiling center of the magnetic field lines. The length scale of the arrows is adapted to improve readability. The absolute value $B_{xz} = \sqrt{B_x^2 + B_z^2}$ ranges from $\approx 60 \G$ in the network fields at $z=1\Mm$ to $\approx 10 \G$ higher up in the atmosphere at about $z=4\Mm$, the upper end of the arrow positions. The animation can be found in the \href{https://iopscience.iop.org/article/10.3847/2041-8213/ae9607}{published journal version} of the article. 
\label{fig:fig-slicing-feature-c}}
\end{figure*}

It can be seen that all three features show field lines that are coiled around a central axis. See for example the magnetic field line sets of constant color belonging to feature A in panel (a) or of feature C (panel i). The central axis itself is however not a straight line but can have writhe. We refer to such features, with magnetic field lines in some areas winding around a common axis, as a \ac{tfr} \citep[see, \eg,][]{1989ApJ...344.1010P}. It turns out that what appears as the \ac{opi} in the magnetogram at $1 \Mm$ (Fig.~\ref{fig:fig1-magnetogram-and-stokes-V}, panel c) or in the \ac{tcp} map (Fig.~\ref{fig:fig1-magnetogram-and-stokes-V}, panel b) is a part of the coiled structure that originates in the magnetic network element and wraps around the inner twisted magnetic field lines. This can also be seen in the movies attached in App.~\ref{app:animations_of_features_a_b_c} that present a $360^{\circ}$ view around the features, and in Fig.~\ref{fig:fig-slicing-feature-c}, where we present a vertical cut through feature C. The magnetic field direction in the plane that is approximately orthogonal to the twist axis indicates a common rotation axis, similar to a \ac{tfr}.

The magnetic field lines close to the coiling center (the ones with constant color) seem to connect from one network polarity to the other (see \eg, features A and B, panels a and d in Fig.~\ref{fig:fig3-3d-rendering-of-field-lines}). The magnetic field lines associated with the \ac{opi} may either connect similarly to the opposite polarity, or they connect to smaller internetwork fields as suggested by feature B, panel (f). Interestingly, the magnetic field lines of feature C that are associated with the \ac{opi} split up into two sets, one set directed to the boundary of the simulation box (likely connecting to internetwork polarities opposite to the simulation box due to the horizontal periodic boundary conditions). The other set points towards the field lines associated with feature B, suggesting the two features may partly be connected. 

\section{Discussion}

In this section, we discuss the appearance of \acp{tfr} as \acp{opi} and compare with reports of simulated \acp{tfr} in the literature. \acp{tfr} have previously been found to form in numerical models. \citet{2015Natur.522..188A} found that \acp{tfr} may form naturally as a result of an active \ac{ssd}. The simulation of \citet{2019ApJ...878...40M}, which is computed with the Bifrost code \citep{2011A&A...531A.154G}, includes twisted magnetic fields in the lower atmosphere, which the authors explain partly by the dynamic velocities in the model. \citet{2021A&A...656L...7C} used a MURaM simulation and found a \ac{tfr} that formed in their simulation that may be related to solar campfires. \citet{2023A&A...673A..79R} found that \acp{tfr} may be built up through component reconnection driven by flows in the convection zone. \citet{2025ApJ...984L..37A} used photospheric magnetograms observed with Hinode \citep[][]{2008ApJ...672.1237L} together with field extrapolations to demonstrate that \acp{tfr} can exist ubiquitously in the \ac{qs}. Recently \citet{2025ApJ...995..137M} simulated the emergence of a bipolar magnetic flux tube and observed the formation of \acp{tfr} in a simulation based on the model by \citet{2019A&A...626A..33H}. In the work by \citet{2025arXiv251220716V}, the authors studied the formation of a \ac{tfr} as a result of a sheared arcade configuration. Qualitatively, the \acp{tfr} in our model compare to \citet[][Fig.~5]{2025arXiv251220716V}. There, the different magnetic field line sets of one footpoint are concentrated in one location, while the other footpoints may connect to multiple locations. In \citet[][Fig.1]{2026A&A...705A..86N} the evolution of a magnetic tornado was presented, which qualitatively looks similar to the magnetic field configuration of feature C. Another possibility for the appearance of the \ac{opi} in the \ac{tcp} might be the presence of sheared arcades. The three features presented in this study show a clearly twisted magnetic field around a common central axis, which speaks against this interpretation. This does not mean, however, that there are no instances of sheared arcades in the simulation. The simulated features might often be a combination of shear and twist as in \citet{2025arXiv251220716V}. To better understand this, a detailed study of the evolution of such features is required in the future.

\begin{figure*}[ht!]
\includegraphics[width=\textwidth,clip]{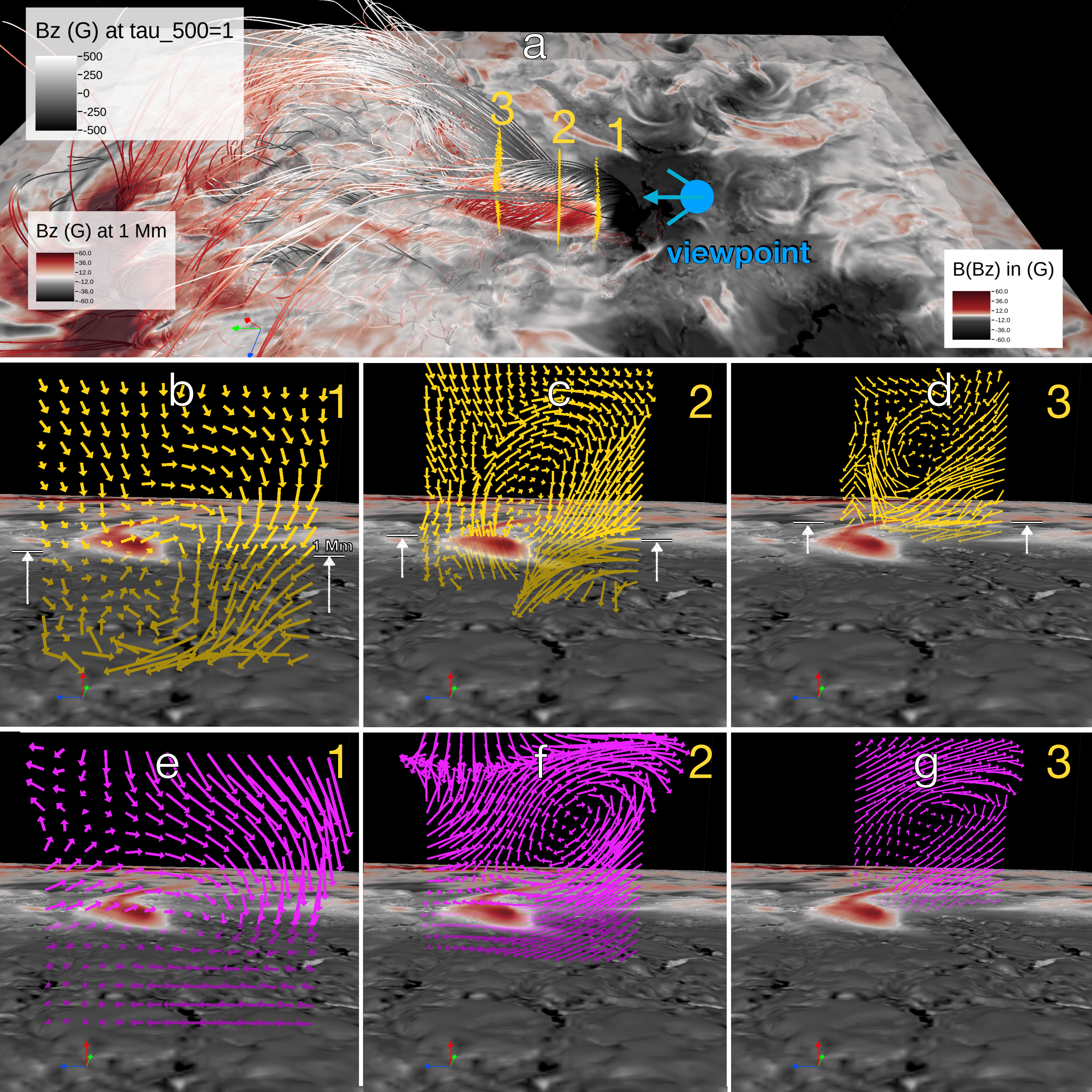}
\caption{Connection between coiled magnetic field lines and vortical motions. Panel (a) shows an overview of the magnetic field lines belonging to feature B. The magnetic field lines are color-coded by the value of the vertical magnetic field component. In the background, the vertical magnetic field at the $\tau_{500}=1$ surface and at $z=1 \Mm$ is shown. Three vertical cuts are indicated, numbered (1), (2), and (3). In panels (b, c, and d), we show the direction of the magnetic field in the planes indicated in panel (a) with yellow arrows. The approximate position of the viewpoint for panels (b--g) is indicated in panel (a). The position of the $z=1 \Mm$ is indicated for reference in panels (b, c, and d). Panels (e, f, and g) are similar to panels (b,c, and d) but show the direction of the velocities with pink arrows. The length scales of the arrows are chosen to improve readability, similar to Fig.~\ref{fig:fig-slicing-feature-c}. The maximum absolute values of the velocities in the planes range from $\approx4\kms$ (position 1) to $\approx7\kms$ (position 3).
\label{fig:feature_b_velocity_and_magnetic_field_vectors}}
\end{figure*}

The twisted magnetic field invites investigating the local velocity field next to such features. Indeed, vortex motions are common in simulations with magnetic features. For example, \citet{2020ApJ...898..137S} and \citet{2021A&A...645A...3Y}, and references therein, demonstrated the presence of strong vortex motions within small magnetic elements in solar plage regions. \citet{2023A&A...675A..94B} studied a simulation showing swirls along a coronal loop and found that they play an important role in the energy transport and structuring of the chromosphere and low corona. We therefore checked whether there are indications of vortex motions inside the \acp{tfr} found in the simulation presented here. In Fig.~\ref{fig:movie_bz} we show the evolution of the vertical magnetic field and the vertical velocity at $1\Mm$. By looking at the evolution of feature B (panelc c and d), it can be seen that, for example at $t=2 \min$, close to the footpoints of the feature, elongated structures of up and downflows are found close to each other. This suggests that the \acp{tfr} might indeed be shaped, to some extent, by vortex motions within the magnetic elements in the simulation. To further illustrate this idea, we show in Fig.~\ref{fig:feature_b_velocity_and_magnetic_field_vectors} three vertical cuts through the magnetic field lines belonging to feature B (see panel a). We present at the respective positions the directions of the magnetic field and velocity in the plane that is approximately perpendicular to the magnetic field lines. We used a Gaussian filter to remove small-scale structures in the velocities. In panels (b, c, and d), it can be seen that in all three cuts, the magnetic field is twisted around a common axis surrounded by a background magnetic field. At the same positions, we show in panels (e,f, and g) the direction of the velocity in the same plane, which indicates a rotational flow around a central axis. The rotational axis of the velocities can be roughly aligned (positions 1 and 3 in panels b, d, e, and g) or slightly off-center (position 2, panels c and f). A twisted magnetic field in the presence of vortex motions might also explain why such features can be seen in regions where the network fields are mostly unipolar. For further understanding of the connection between the rotational motions and the twisted magnetic field, a more detailed study is required, which includes the evolution of such features.

\citet{2017ApJ...848...38I} presented the evolution of a chromospheric jet that appeared similar to observed spicules. For their studies, the authors simulated a \ac{qs} region of $9\Mm \times 9 \Mm$ horizontal extent with a unipolar magnetic field of $10 \G$ and found that vortex motions in the lower atmosphere can lead to a twisted magnetic field. In their simulation, the velocity streamlines of the vortex and the accompanying twisted magnetic field are slightly inclined from the vertical axis. In our simulation the vortex motions as well as the \ac{tfr} are part of the loops connecting between the network polarities and thus show even higher inclination from the vertical axis. \citet{2017ApJ...848...38I} describe a mechanism by which such a magnetic field configuration can lead to a chromospheric jet due to the Lorentz force. The idea of a rotating flow in the presence of the \acp{tfr} is also consistent with the results of \citet[][Fig.~7]{2026A&A...706A..53C}, who found cospatial occurrence of \acp{rbe} and \acp{rre} in the same simulation as presented here. 

\citet{2023MNRAS.523..974M} used the simulation of \citet{2017ApJ...848...38I} and forward modeled the Stokes signals of the \ion{Ca}{ii} $849.8 \mathrm{\,nm}$ and the \ion{Fe}{i} $846.8 \mathrm{\,nm}$ lines. By doing so, the authors could show that the chromospheric jets show imprints in the forward-modeled linear polarization. The forward-modeled total linear polarization (TLP, not shown) from our simulation shows some signal at the location of feature C. Features A and B are, however, hardly visible in the TLP.

\section{Summary and conclusions}

Novel high-resolution and high-precision observations from the \ac{scip} instrument onboard the third flight of the Sunrise observatory revealed fine-structured \acfp{opi} in the magnetic network.
The aim of this work is to identify whether a numerical model of the solar atmosphere computed with the \ac{muramche} code can produce similar fine-structured magnetic network fields and to identify the underlying magnetic field topology.

By forward modeling the Stokes-$V$ signal of the \caefft line, we found that such features are rather common in the simulation and appear frequently around the borders of the network polarities in the lower to middle chromosphere at roughly $1\Mm$. We selected three features and studied the 3D magnetic field configuration. We found that what appears as an \ac{opi} in the magnetogram and the Stokes-$V$ signal is in all three cases associated with a twist of the magnetic field similar to a \acf{tfr}. Having a footpoint in one network polarity, the magnetic field lines either connect in a coiled way to the opposite network polarity or connect to the internetwork magnetic field. The latter seems to happen if the magnetic field lines, due to the winding, reach the lower-lying internetwork magnetic field polarities. We find that some of the features, such as features A and B (see Fig.~\ref{fig:fig1-magnetogram-and-stokes-V}), appear throughout the whole presented time series of $\approx 21 \min$ as \acp{opi} in the \ac{tcp} map and the magnetogram at $1\Mm$.

Our work demonstrates that \acp{tfr} form naturally in the \ac{muramche} model at a relatively moderate horizontal resolution of $23.4\km$, likely as a result of vortex flows in the magnetic elements. Whether they appear as \acp{opi} in the magnetogram at $z=1\Mm$ or in the forward-modeled \ac{tcp} map depends on their evolution and geometry. This means if the twisted field lines are oriented such that they overlap with the formation height region of the \caefft Stokes-V signal, they become visible in the \ac{tcp} map. The \acp{opi} form ubiquitously around the network patches, which is consistent with the location of vortices in the simulation of \citet[][Fig.~4]{2021A&A...645A...3Y}. Their presence and their interaction with each other may add to the dynamics in the chromosphere and corona of this model. Recently \citet{2024A&A...692A...6O} used the same model and found that the dynamics in the chromosphere lead to an improved match of the forward-modeled \mghk lines with observations. The corrugated transition region discussed in \citet{2024A&A...692A...6O} might partly be a result of the dynamic \acp{tfr}. In addition, dynamic H$\alpha$ features were identified in the model, including \acp{rre} and \acp{rbe} \citep{2025A&A...701A.294C} and spicules \citep{2026A&A...706A..53C}. Whether there is a correlation between the \acp{tfr} and the spicule events needs to be shown in future work. We conclude that by forward modeling of the \caefft Stokes-$V$ profiles, our model provides evidence that the \textsc{Sunrise~III}/\ac{scip} observations of \citet{2026ApJ..1008L...9K} show signatures of \acp{tfr} in the observed \ac{tcp} of the chromosphere, although other explanations cannot be ruled out.

\begin{acknowledgments}
\textsc{Sunrise~iii} is supported by funding from the Max-Planck-Förderstiftung (Max Planck Foundation), NASA under Grants \#80NSSC18K0934 and \#80NSSC24M0024 (“Heliophysics Low Cost Access to Space” program), and the ISAS/JAXA Small Mission-of-Opportunity program and JSPS KAKENHI Grant Numbers JP18H05234 and JP23K25916. This research has received financial support from the European Union’s Horizon 2020 research and innovation program under grant agreement No. 824135 (SOLARNET) and No. 101097844 (WINSUN) from the European Research Council (ERC). It has also been funded by the Deutsches Zentrum für Luft- und Raumfahrt e.V. (DLR, grant no. 50 OO 1608). The Spanish contributions have been funded by the Spanish MCIN/AEI/10.13039/501100011033 under projects RTI2018-096886-B-C5, PID2021-125325OB-C5, and PID2024-156066OB-C5, and from "Center of Excellence Severo Ochoa" awards to IAA-CSIC (SEV-2017-0709, CEX2021-001131-S), all co-funded by "ERDF A way of making Europe".
\end{acknowledgments}

\appendix

\section{Computation of the total circular polarization
}
\label{app:tcp}

We compute the \ac{tcp} as follows \citep[see also,][]{2026ApJ..1008L...9K}:

\begin{equation}
    \text{TCP} = \frac{\int_{\lambda_0-\lambda_1}^{\lambda_0-\lambda_{\mathrm{c}}} \frac{V(\lambda)}{I(\lambda)} \mathrm{d}\lambda - \int^{\lambda_0+\lambda_1}_{\lambda_0+\lambda_{\mathrm{c}}} \frac{V(\lambda)}{I(\lambda)} \mathrm{d}\lambda}{\int_{\lambda_0-\lambda_1}^{\lambda_0-\lambda_{\mathrm{c}}}  \mathrm{d}\lambda + \int^{\lambda_0+\lambda_1}_{\lambda_0+\lambda_{\mathrm{c}}} \mathrm{d}\lambda}
\end{equation}

Here $\lambda_0$ denotes the line-center wavelength of the \caefft line. We follow \citet{2026ApJ..1008L...9K} and chose $\lambda_1 = 36.5 \ppm$ and $\lambda_\mathrm{c} = 3.95 \ppm$. The adopted values of $\lambda_1$ and $\lambda_\mathrm{c}$ are based on the \ac{scip} instrumental specifications. We checked an alternative definition of the \ac{tcp} where we integrated the \ac{tcp} up to $\lambda_0$, that is with $\lambda_\mathrm{c}=0$, and found no significant difference in the appearance of the \ac{tcp} map.

\section{Animations of the viewing angles of features A, B, and C}
\label{app:animations_of_features_a_b_c}
\begin{figure*}[ht!]
\begin{interactive}{animation}{animation_figure_7}
\includegraphics[width=\textwidth,clip]{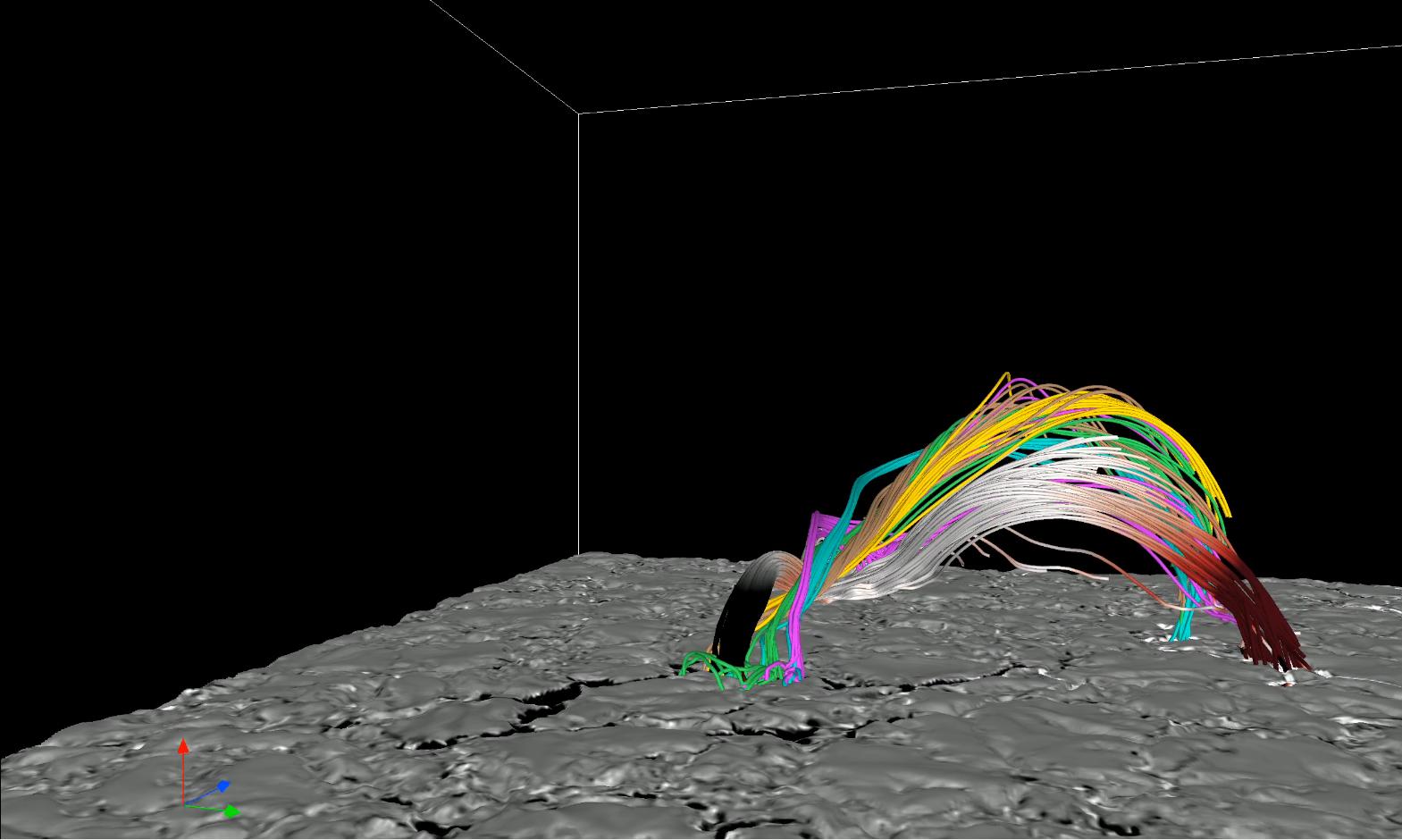}
\end{interactive}\caption{Animation showing a $360^{\circ}$ view of feature A as an addition to Fig.~\ref{fig:fig3-3d-rendering-of-field-lines}. The duration of the animation is $24 \sec$. The example frame shows the coiled magnetic field lines from the side. The animation can be found in the \href{https://iopscience.iop.org/article/10.3847/2041-8213/ae9607}{published journal version} of the article.\label{fig:animation_feature_a}}
\end{figure*}

\begin{figure*}[ht!]
\begin{interactive}{animation}{animation_figure_8.mp4}
\includegraphics[width=\textwidth,clip]{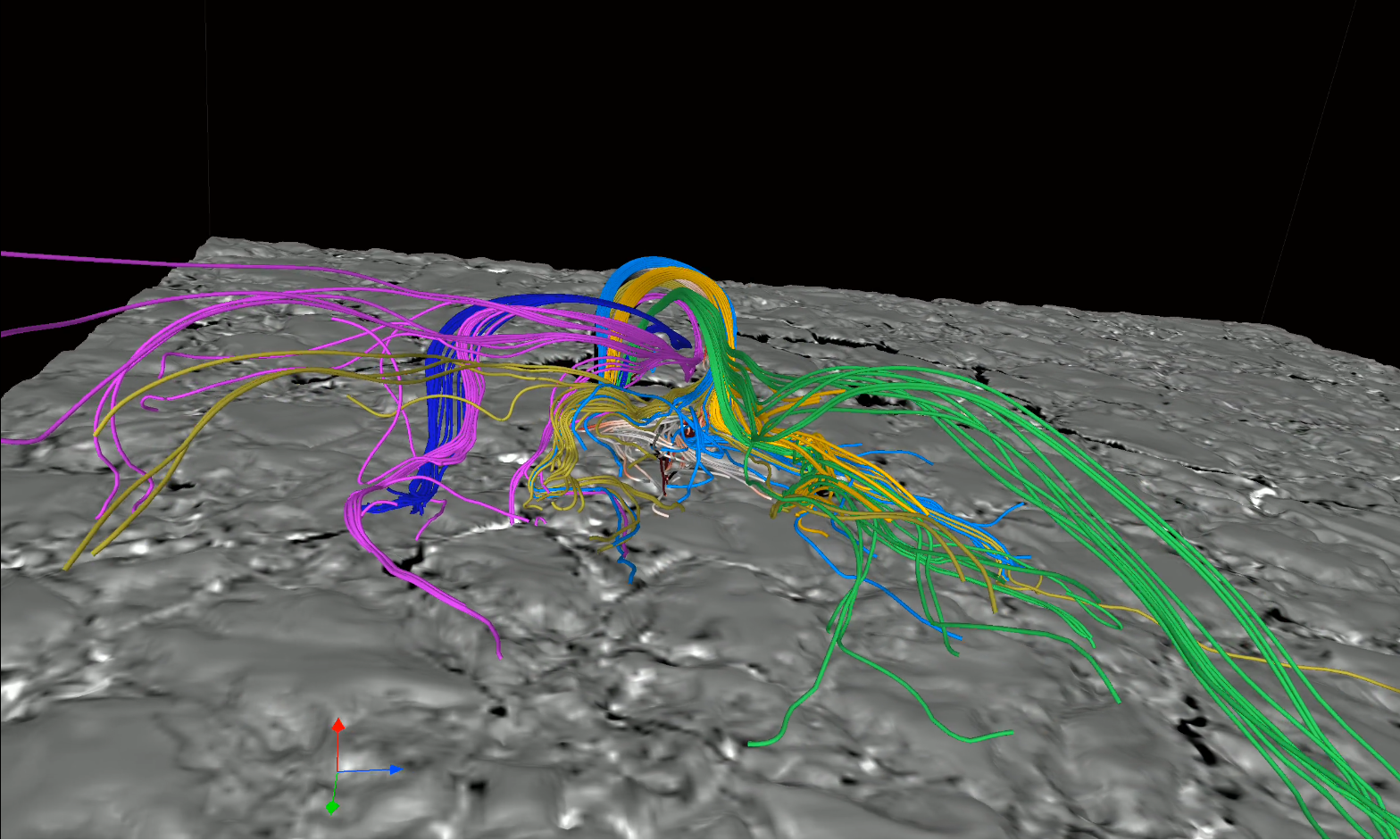}
\end{interactive}\caption{Animation showing a $360^{\circ}$ view of feature B as an addition to Fig.~\ref{fig:fig3-3d-rendering-of-field-lines}. The example frame shows the coiled magnetic field lines from a viewpoint oriented approximately along the central axis of the twist. The duration of the animation is $24 \sec$. The animation can be found in the \href{https://iopscience.iop.org/article/10.3847/2041-8213/ae9607}{published journal version} of the article.
\label{fig:animation_feature_b}}
\end{figure*}

\begin{figure*}[ht!]
\begin{interactive}{animation}{animation_figure_9.mp4}
\includegraphics[width=\textwidth,clip]{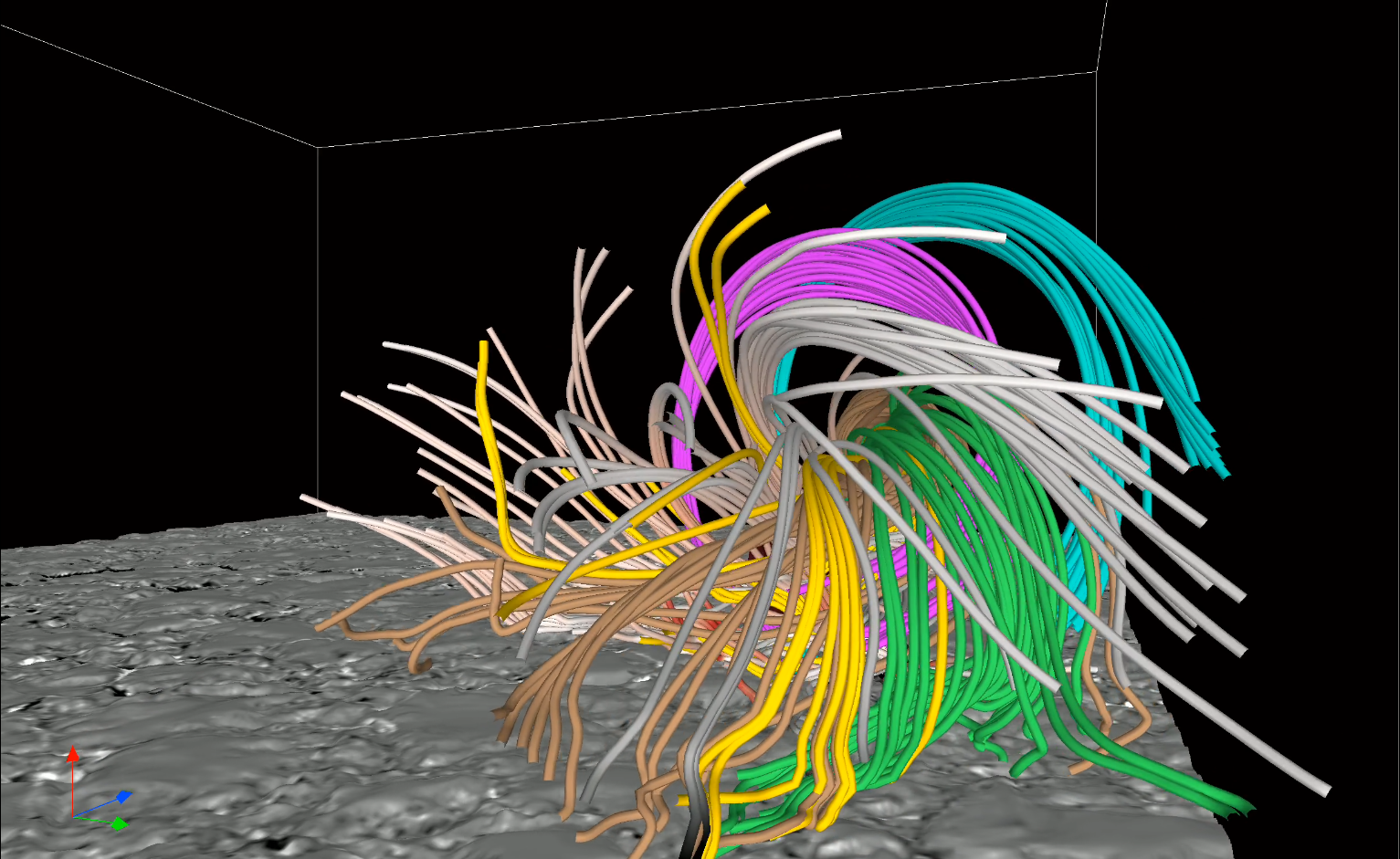}
\end{interactive}\caption{Animation showing a $360^{\circ}$ view of feature C as an addition to Fig.~\ref{fig:fig3-3d-rendering-of-field-lines}. The example frame shows the coiled magnetic field lines from a viewpoint oriented approximately along the central axis of the twist. The duration of the animation is $24 \sec$. The animation can be found in the \href{https://iopscience.iop.org/article/10.3847/2041-8213/ae9607}{published journal version} of the article.
\label{fig:animation_feature_c}}
\end{figure*}

In this appendix, we present animations of features A (Fig.~\ref{fig:animation_feature_a}), B (Fig.~\ref{fig:animation_feature_b}), and C (Fig.~\ref{fig:animation_feature_c}) to provide a $360^{\circ}$ view of the magnetic field lines.

\bibliography{bib}{}
\bibliographystyle{aasjournalv7}

\end{document}